\documentclass[12pt]{article}

\usepackage{graphicx}
\begin{document}

\begin{center}
{\bf Magnetically charged AdS black holes and Joule--Thomson expansion } \\
\vspace{5mm} S. I. Kruglov
\footnote{E-mail: serguei.krouglov@utoronto.ca}
\underline{}
\vspace{3mm}

\textit{Department of Physics, University of Toronto, \\60 St. Georges St.,
Toronto, ON M5S 1A7, Canada\\
Department of Chemical and Physical Sciences, University of Toronto,\\
3359 Mississauga Road North, Mississauga, Ontario L5L 1C6, Canada}, \\
Canadian Quantum Research Center, 204-3002 32 Ave Vernon, BC V1T 2L7, Canada\\
\vspace{5mm}
\end{center}
\begin{abstract}
The process of the Joule--Thomson adiabatic expansion within RNED-AdS spacetime is investigated. The isenthalpic $P-T$ diagrams and the inversion temperature were depicted. The inversion temperature depends on the magnetic charge and RNED coupling constant of black holes. When
the Joule--Thomson coefficient vanishes, cooling-heating phase transition occurs. We consider the cosmological constant as a thermodynamic pressure and the black hole mass is treated as the chemical enthalpy.
 \end{abstract}

General relativity possesses black hole solutions which are characterized by the mass, angular momentum and charges.
The authors of \cite{Bardeen}, \cite{Jacobson}, \cite{Padmanabhan}, \cite{Hawking} discovered that a black hole is a thermodynamic system with entropy and temperature and the first law of black hole thermodynamics was formulated. But in that form of black hole thermodynamics the $P-V$ term of ordinary thermodynamics was absent. Later, this problem was solved by adding the negative cosmological constant treated as a thermodynamic pressure which is conjugated to a volume \cite{Kubiznak_1}, \cite{Kastor}, \cite{Dolan_1}.
Thus, we need to consider anti-de Sitter (AdS) spacetime to include $P-V$ term in the first law of black hole thermodynamics. It was shown that in such extended theory of gravity (Einstein-AdS) phase transitions in black holes occur \cite{Page}. Such black holes phase transitions are similar to liquid-gas (for Van der Waals gases) phase transitions in ordinary thermodynamics \cite{Dolan}, \cite{Kubiznak}, \cite{Mann}, \cite{Teo}, \cite{Dutta}, \cite{Farsam}, \cite{Mann_1}, \cite{Zou}, \cite{Altamirano}, \cite{Cai}, \cite{Mo}, \cite{Altamirano_1}, \cite{Hansen}, \cite{Rajagopal}. Within Einstein-AdS theory of gravity a holographic principle was detected with the connection to conformal field theories \cite{Maldacena}, \cite{Witten}, \cite{Witten1}.

In this letter, we study the Joule--Thomson adiabatic thermal expansion, with heating and cooling regimes, in the framework of rational nonlinear electrodynamics (RNED) AdS black holes. The Joule--Thomson adiabatic expansion happens when a black hole mass, which is an enthalpy, is constant during the expansion. When the pressure of the expanding black holes decreases then at some point (the inversion pressure) a cooling-heating transition happens \cite{Aydinir}, \cite{Yaraie}, \cite{Mo_1}, \cite{Rizwan}. RNED was proposed in \cite{Kr0}, \cite{Kr2} to remove singularities and to take into consideration quantum gravity corrections to classical Maxwell fields. The model of RNED at weak-field limit becomes the Maxwell electrodynamics and coupled to gravity perfectly describes the universe inflation \cite{Kr} and gives the correct shadow of M87* black hole \cite{Kr1}. Thermodynamics of RNED-AdS magnetic black holes in the extended phase space was studied in \cite{Kr3}.

Einstein–RNED  AdS space-time is described by the action
\begin{equation}
I=\int d^{4}x\sqrt{-g}\left(\frac{R-2\Lambda}{16\pi G_N}+\mathcal{L}(\mathcal{F}) \right),
\label{1}
\end{equation}
where $G_N$ is the Newton constant.
The negative cosmological constant can be written as $\Lambda=-3/l^2$, where $l$ is the AdS radius. We use RNED Lagrangian \cite{Kr0}
\begin{equation}
{\cal L}(\mathcal{F}) =-\frac{{\cal F}}{4\pi(1+2\beta{\cal F})},
\label{2}
\end{equation}
where  ${\cal F}=F^{\mu\nu}F_{\mu\nu}/4=(B^2-E^2)/2$, $F_{\mu\nu}=\partial_\mu A_\nu-\partial_\nu A_\mu$ is the field strength. It is worth mentioning that the singularity of the electric field in the center of point-like charges is absent and the maximum electric field is $E(0)=1/\sqrt{\beta}$ \cite{Kr0}. Setting $\beta=0$ in Eqs. (1) and (2), one finds the action for the Einstein--AdS space-time.
From action (1) we obtain the field equations
\begin{equation}
R_{\mu\nu}-\frac{1}{2}g_{\mu \nu}R+\Lambda g_{\mu \nu} =8\pi G_N T_{\mu \nu},
\label{3}
 \end{equation}
\begin{equation}
\partial _{\mu }\left( \sqrt{-g}\mathcal{L}_{\mathcal{F}}F^{\mu \nu}\right)=0,
\label{4}
\end{equation}
where $\mathcal{L}_{\mathcal{F}}=\partial \mathcal{L}/\partial \mathcal{F}$.
Here, we use the four-dimensional static spherical symmetry line element squared
\begin{equation}
ds^{2}=-f(r)dt^{2}+\frac{1}{f(r)}dr^{2}+r^{2}\left( d\theta
^{2}+\sin ^{2}\theta d\phi ^{2}\right).
\label{5}
\end{equation}
We study magnetic black holes because electrically charged black holes possess singularity \cite{Bronnikov}.  The metric function with the spherical symmetry can be written in the form
\begin{equation}
f(r)=1-\frac{2m(r)G_N}{r}.
\label{6}
\end{equation}
The mass function is given by
\[
m(r)=m_0+\frac{q_m^{3/2}}{8\sqrt{2}\beta^{1/4}}\biggl[\ln\left(\frac{r^2-\sqrt{2q_m}\beta^{1/4}r+q_m\sqrt{\beta}}{r^2
+\sqrt{2q_m}\beta^{1/4}r+q_m\sqrt{\beta}}\right)
\]
\begin{equation}
+2\arctan\left(\frac{\sqrt{q_m}\beta^{1/4}+\sqrt{2}r}{\sqrt{q_m}\beta^{1/4}}\right)
-2\arctan\left(\frac{\sqrt{q_m}\beta^{1/4}-\sqrt{2}r}{\sqrt{q_m}\beta^{1/4}}\right)\biggr]-\frac{r^3}{2G_Nl^2},
\label{7}
\end{equation}
where $m_0$ is the Schwarzschild mass which is the constant of integration. When the cosmological constant is zero, $\Lambda=0$ ($l=\infty$), one comes to the mass function given in \cite{Kr2}. The magnetic mass of a black hole is finite and reads as
\begin{equation}
m_M=4\pi\int_0^\infty \frac{q_m^2}{8\pi(r^4+\beta q_m^2)}r^2dr=\frac{\pi q_m^{3/2}}{4\sqrt{2}\beta^{1/4}}
\approx 0.56\frac{q_m^{3/2}}{\beta^{1/4}}.
\label{8}
\end{equation}
It should be noted that in the Maxwell case, $\beta=0$, the magnetic energy is infinite.
With the help of Eqs. (6) and (7) one obtains the metric function
\begin{equation}
f(r)=1-\frac{2m_0G_N}{r}-\frac{q_m^{3/2}G_Ng(r)}{4\sqrt{2}\beta^{1/4}r}+\frac{r^2}{l^2},
\label{9}
\end{equation}
with
\[
g(r)\equiv \ln\left(\frac{r^2-\sqrt{2q_m}\beta^{1/4}r+q_m\sqrt{\beta}}{r^2+\sqrt{2q_m}\beta^{1/4}r+q_m\sqrt{\beta}}\right)
\]
\[
+2\arctan\left(\frac{\sqrt{q_m}\beta^{1/4}+\sqrt{2}r}{\sqrt{q_m}\beta^{1/4}}\right)
-2\arctan\left(\frac{\sqrt{q_m}\beta^{1/4}-\sqrt{2}r}{\sqrt{q_m}\beta^{1/4}}\right).
\]
In the following we use the units with $G_N=1$. Making use of Eq. (9) and the equation for the horizon radius $r_+$, $f(r_+)=0$, we obtain the black hole mass
\begin{equation}
M\equiv m_0+m_M=\frac{r_+}{2}+\frac{r_+^3}{2l^2}+\frac{\pi q_m^{3/2}}{4\sqrt{2}\beta^{1/4}}-\frac{q_m^{3/2}g(r_+)}{8\sqrt{2}\beta^{1/4}}.
\label{10}
\end{equation}
The Hawking temperature $T=f'(r)|_{r=r_+}/(4\pi)$, where $f'(r)=\partial f(r)/\partial r$, becomes \cite{Kr2}
\begin{equation}
T=\frac{1}{4\pi}\biggl(\frac{1}{r_+}+\frac{3r_+}{l^2}-\frac{q_m^2r_+}{r_+^4+\beta q_m^2}\biggr).
\label{11}
\end{equation}
At $\beta=0$ in Eq. (11), we arrive at the Hawking temperature of Maxwell-AdS black holes.
Making use of the equation for the black hole pressure $P=3/(8\pi l^2)$ and Eq. (11), one finds the equation of state
\begin{equation}
P=\frac{T}{2r_+}-\frac{1}{8\pi r_+^2}+\frac{q_m^2}{8\pi(r_+^4+\beta q_m^2)}.
\label{12}
\end{equation}

It should be noted that first law of black hole thermodynamics states that the black hole mass $M$ is the enthalpy.
The Joule--Thomson adiabatic process is isenthalpic expansion and occurs at the constant black hole mass. To study cooling-heating phases, one introduces the Joule--Thomson thermodynamic coefficient
\begin{equation}
\mu_J=\left(\frac{\partial T}{\partial P}\right)_M=\frac{1}{C_P}\left[ T\left(\frac{\partial V}{\partial T}\right)_P-V\right]=\frac{(\partial T/\partial r_+)_M}{(\partial P/\partial r_+)_M}.
\label{13}
\end{equation}
Equation (13) shows that the Joule--Thomson coefficient $\mu_J$ is the slope in $P-T$ diagrams at the constant mass.
At the inversion temperature $T_i$ (when $\mu_J(T_i)=0$) the sign of the coefficient $\mu_J$ is changed. When the initial temperature during the black hole expansion is bigger than inversion temperature $T_i$ ($\mu_J>0$), the temperature decreases which corresponds to the cooling phase. For the heating phase, the initial temperature is lower than $T_i$ and the temperature increases. Making use of Eq. (13) and equation $\mu_J(T_i)=0$ we find the inversion temperature
\begin{equation}
T_i=V\left(\frac{\partial T}{\partial V}\right)_P=\frac{r_+}{3}\left(\frac{\partial T}{\partial r_+}\right)_P,
\label{14}
\end{equation}
where we use the expression for black hole thermodynamic volume $V=4\pi r_+^3/3$. The inversion temperature corresponds to the temperature maximum in the $P-T$ diagram and is a borderline between cooling and heating processes. The inversion temperature
line connects points in maxima of $P-T$ diagrams at the constant black hole mass and separates cooling and heating phases \cite{Yaraie}, \cite{Mo}. The black hole equation of state (11) can be represented in the form
\begin{equation}
T=\frac{1}{4\pi r_+}+2P r_+-\frac{q_m^2r_+}{4\pi(r_+^4+q_m^2\beta)}.
\label{15}
\end{equation}
Setting $\beta=0$ in Eq. (15) we find the equation of state for Maxwell-AdS black holes.
From Eq. (10) one obtains the equation for pressure
\begin{equation}
P=\frac{3}{4\pi r_+^3}\left[M-\frac{r_+}{2}-\frac{\pi q_m^{3/2}}{4\sqrt{2}\beta^{1/4}}+\frac{q_m^{3/2}g(r_+)}{8\sqrt{2}\beta^{1/4}}\right].
\label{16}
\end{equation}
Making use of Eqs. (15) and (16) we depicted the $P-T$ diagrams in Fig. 1.
\begin{figure}[h]
\includegraphics {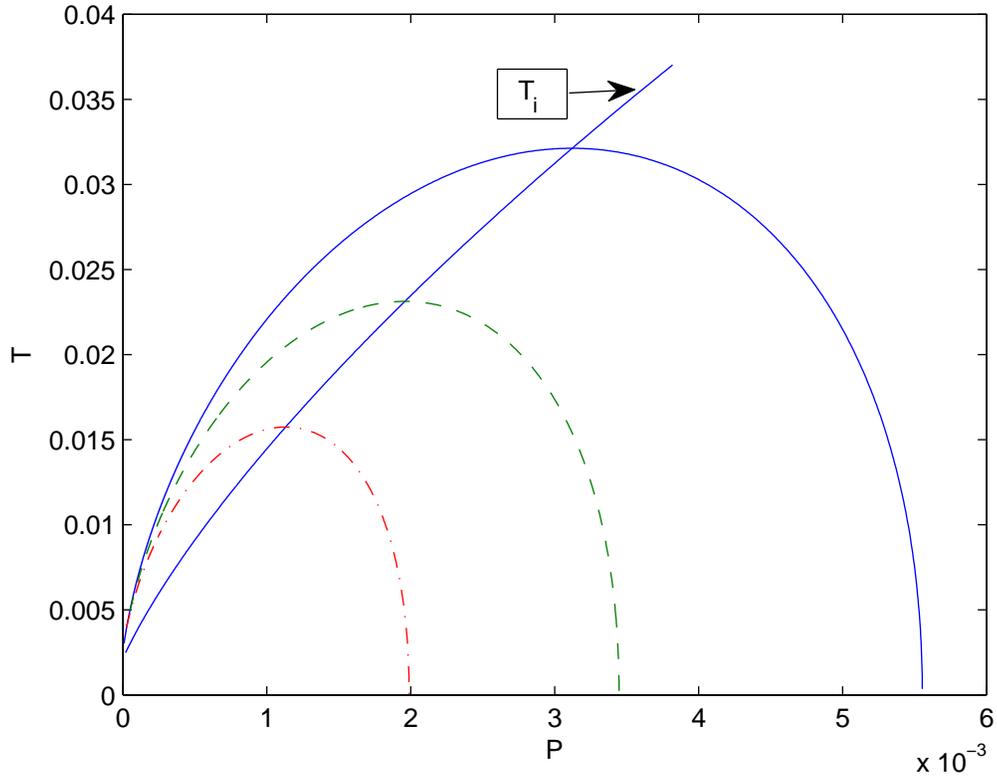}
\caption{\label{fig.1} The plots of the temperature $T$ versus pressure $P$  and the inversion temperature $T_i$ with $q_m=10$, $\beta=1$.
The dashed-dotted line corresponds to mass $M=13$, the dashed line corresponds to $M=14$, and the solid line corresponds to $M=15$.}
\end{figure}
By using Eqs. (14) and (15) one finds the equation for the inversion pressure
\begin{equation}
P_i=\frac{q_m^2\left(3r_+^4+q_m^2\beta\right)}{8\pi \left(r_+^4+q_m^2\beta\right)^2}-\frac{1}{4\pi r_+^2}.
\label{17}
\end{equation}
Substituting $P_i$ from Eq. (17) into Eq. (15) we obtain the inversion temperature
\begin{equation}
T_i=\frac{q_m^2r_+^5}{2\pi \left(r_+^4+q_m^2\beta\right)^2}-\frac{1}{4\pi r_+}.
\label{18}
\end{equation}
Numerical solutions to equation $P_i=0$, (17),  and the minimum of the inversion temperature (18) at $q=1$ are given in Table 1.
\begin{table}[ht]
\caption{The minimum of the event horizon radius corresponding to the minimum of the inversion temperature at $q=1$}
\centering
\begin{tabular}{c c c c c c c c c c c}\\[1ex]
\hline
$\beta$ & 0.01 & 0.03 & 0.05 & 0.07 & 0.09 & 0.1 & 0.2 & 0.3   \\[0.5ex]
\hline
$r_i^{min}$ & 1.220 & 1.211 & 1.201 & 1.190 & 1.179 & 0.174 & 1.103 & 0.937  \\[0.5ex]
\hline
$T_i^{min}$ & 0.0216 & 0.0215 & 0.0214 & 0.0213 & 0.0212 & 0.0210 & 0.0199 & 0.0153  \\[0.5ex]
\hline
\end{tabular}
\end{table}
We did not present solutions for negative values of the minimum of inversion temperature which are non-physical.
Table 1 shows that when coupling $\beta$ increases the minimum of the inversion temperature decreases.
From Eqs. (17) and (18) at $P_i=0$, $\beta=0$, one finds the minimum of the inversion temperature for Maxwell-AdS magnetic black holes
\begin{equation}
T_i^{min}=\frac{1}{6\sqrt{6}\pi q_m}, ~~~~r_+^{min}=\frac{\sqrt{6}q_m}{2}.
\label{19}
\end{equation}
Equation (19) agrees with the result obtained in \cite{Aydinir} for electrically charged Maxwell-AdS black holes.
Taking into account the critical temperature $T_c$ for $\beta=0$ \cite{Kr3} and Eq. (19), we obtain the relation $T_i^{min}=T_c/2$ that holds for electrically charged Maxwell-AdS black holes \cite{Aydinir}. In our case $\beta\neq 0$ one has $T_i^{min}\neq T_c/2$.
Equations (17) and (18) define the inversion temperature $T_i$ versus $P_i$ in the parametric form. The plots of $T$ versus pressure $P$ for various black hole masses and $T_i(P_i)$ are depicted in Fig. 1. According to Fig. 1 the inversion point increases when the black hole mass increases. The plots of the inversion curve $P_i-T_i$ are given in Figs. 2 and 3 for different magnetic charges $q_m$ and couplings $\beta$.
\begin{figure}[h]
\includegraphics {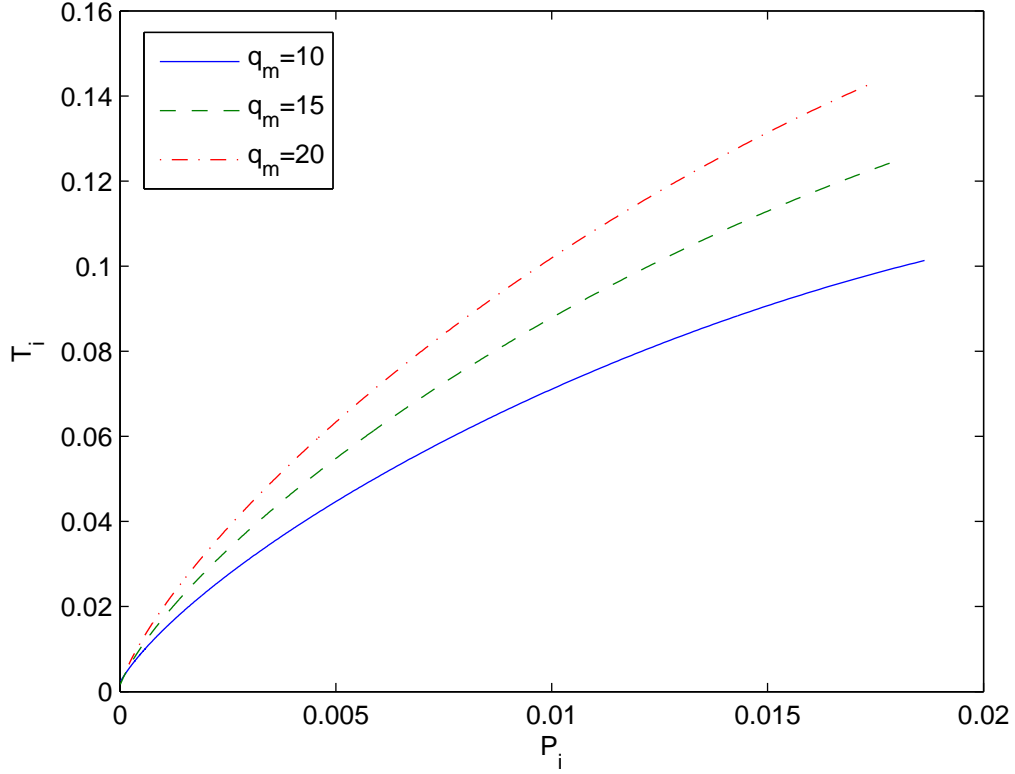}
\caption{\label{fig.2} The plots of the inversion temperature $T_i$ versus pressure $P_i$ for $q_m=10$, $15$ and $20$, $\beta=1$.}
\end{figure}
\begin{figure}[h]
\includegraphics {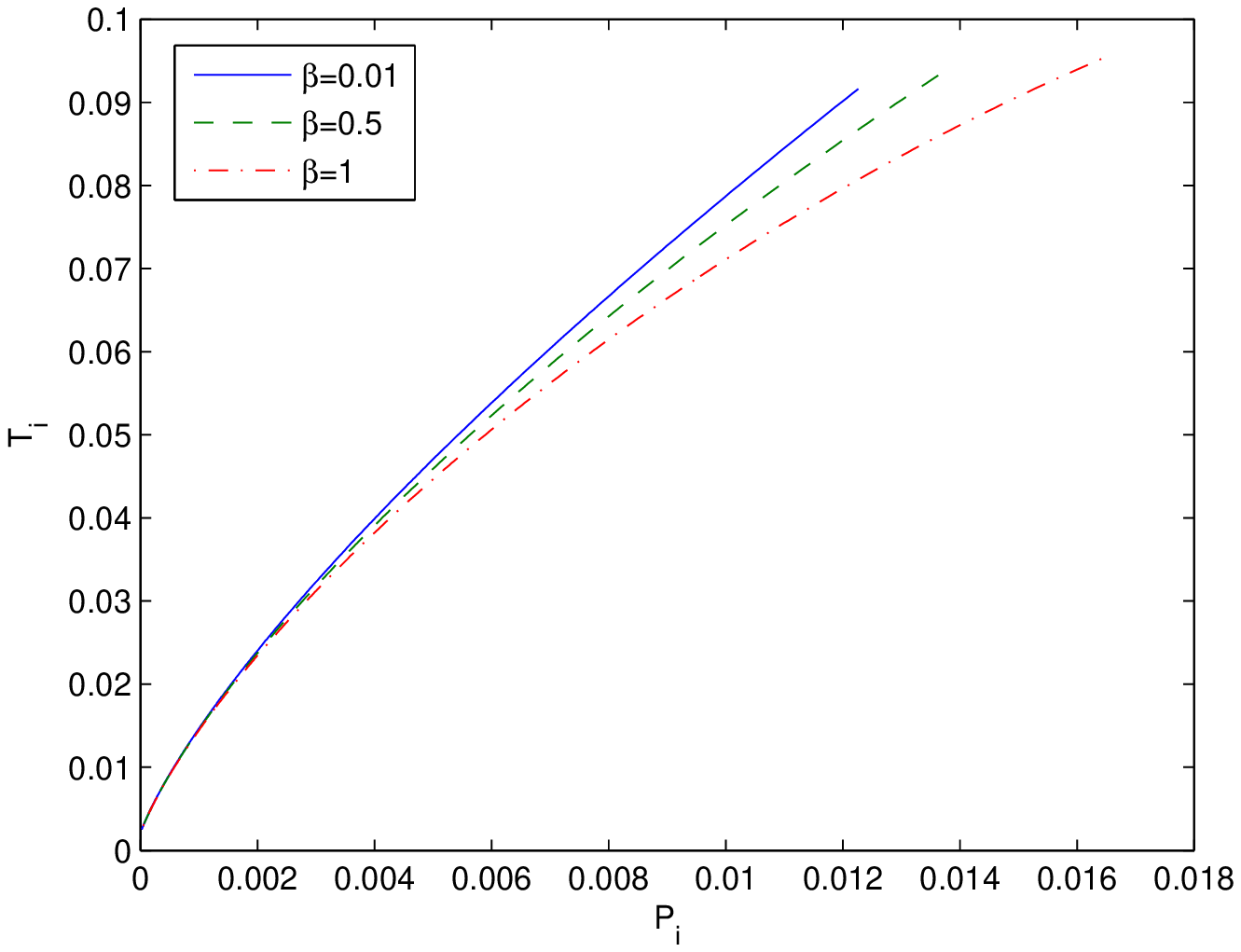}
\caption{\label{fig.3} The plots of the inversion temperature $T_i$ versus pressure $P_i$ for $\beta=0.01$, $0.5$ and $1$, $q_m=10$.}
\end{figure}
In accordance with Fig. 2 when magnetic charge $q_m$ increases, for fixes coupling $\beta$, the inversion temperature increases.
Fig. 3 shows that if the coupling $\beta$ increases at fixed magnetic charge $q_m$, the inversion temperature decreases.
With the aid of Eqs. (15) and (16) we find
\[
\left(\frac{\partial T}{\partial r_+}\right)_M=\frac{1}{2\pi r_+^2}-\frac{3(8\sqrt{2}\beta^{1/4}M-2q_m^{3/2}\pi+q_m^{3/2}g(r_+))}{8\sqrt{2}\pi \beta^{1/4}r_+^3}+\frac{q_m^2(3r_+^4+\beta q_m^2)}{2\pi(r_+^4+\beta q_m^2)^2},
\]
\begin{equation}
\left(\frac{\partial P}{\partial r_+}\right)_M=\frac{3}{4\pi r_+^3}-\frac{9(8\sqrt{2}\beta^{1/4}M-2q_m^{3/2}\pi+q_m^{3/2}g(r_+))}{32\sqrt{2}\pi \beta^{1/4}r_+^4}+\frac{3q_m^2}{8\pi r_+(r_+^4+\beta q_m^2)}.
\label{20}
\end{equation}
Making use of Eqs. (13) and (20) we obtain the Joule--Thomson coefficient
\[
\mu_J(M,r_+)=\frac{2r_+(1-2a+b)}{3(1-a+c)},~~~a=\frac{3(8\sqrt{2}\beta^{1/4}M-2q_m^{3/2}\pi+q_m^{3/2}g(r_+))}{8\sqrt{2}\pi \beta^{1/4}r_+},
\]
\begin{equation}
b=\frac{q_m^2r_+^2(3r_+^4+\beta q_m^2)}{(r_+^4+\beta q_m^2)^2},~~~~~c=\frac{q_m^2r_+^2}{2(r_+^4+\beta q_m^2)}.
\label{21}
\end{equation}
If the Joule--Thomson coefficient is positive for some model parameters, we have a cooling process and when $\mu_J<0$, a heating process occurs. The region  with $\mu_J>0$ in Fig. 1 takes place in the left side of the inversion temperature borderline and the region with $\mu_J<0$ is in the right side of the borderline $T_i$.

In this letter we have investigated cooling and heating phase transitions of RNED-AdS black holes via the Joule--Thomson adiabatic expansion. Isenthalpic $P-T$ diagrams and the inversion temperatures curves for different parameters were depicted in Fig. 1. The plots of the inversion temperature versus the inversion pressure for some magnetic charges and RNED couplings of black holes are given in Figs. 2 and 3.

\end{document}